\documentclass[letterpaper, 10 pt, conference]{ieeeconf}  % Comment this line out if you need a4paper

\IEEEoverridecommandlockouts                     
\usepackage{dsfont}

\usepackage{amsthm}
\usepackage{amssymb}
\usepackage{romannum}
\usepackage{xcolor}
\usepackage{cite}
\usepackage{amsmath,amsfonts}
\allowdisplaybreaks
\usepackage{textcomp}
\usepackage{graphicx} 
\usepackage{algorithm}
\usepackage[capitalize]{cleveref}
\usepackage{algpseudocode}
\usepackage{url}
\usepackage{bbm}
\usepackage{balance}

\renewcommand{\baselinestretch}{0.93}
\usepackage[inkscape=false]{svg}
\usepackage{pgfplots}
\usepackage{pifont}
\pgfplotsset{compat=1.8}
\usepackage{textcomp}

\newtheorem*{assumption*}{Assumption}

\newtheorem{definition}{Definition}

\usepackage{multirow}
\usepackage{overpic}

\usepackage{xcolor}
\usepackage{booktabs}
\usepackage{glossaries}

\newcommand{\tup}[1]{\left(#1\right)}
\newcommand{\set}[1]{\left\{ #1 \right\}}
\newacronym{acr:pt}{PT}{public transport}
\newacronym{acr:tx}{TX}{taxi}

\newacronym{acr:mso}{MSOs}{mobility service operators}
\newcommand{\modeset}{\mathcal{M}}
\newcommand{\modeindex}{m}
\newcommand{\graph}{\Gamma}  
\newcommand{\nodes}{\mathcal{V}}  
\newcommand{\edges}{\mathcal{E}} 
\newcommand{\labelsmap}{\ell}
\newcommand{\labels}{\mathcal{S}}
\newcommand{\labelsingle}{s}
\newcommand{\usergroupset}{\mathcal{N}}
\newcommand{\usergroupindex}{n}
\newcommand{\vot}{\gamma}
\newcommand{\paths}{p}
\newcommand{\pathset}{\mathcal{P}}
\newcommand{\requestset}{\mathcal{R}}
\newcommand{\request}{r}
\newcommand{\requestnumber}{q}

\newcommand{\userset}{\mathcal{I}}
\newcommand{\userindex}{i}

\newcommand{\useractionset}{\mathcal{Y}}
\newcommand{\useraction}{\mathbf{y}}
\newcommand{\userobj}{C}

\newcommand{\flow}{y}

\newcommand{\operatorset}{\mathcal{K}}
\newcommand{\operatorindex}{k}

\newcommand{\operatoractionset}{\mathcal{X}}
\newcommand{\operatoraction}{\mathbf{x}}
\newcommand{\operatorobj}{f}

\newcommand{\ptfre}{q}
\newcommand{\govactionset}{\mathcal{Z}}
\newcommand{\govaction}{\mathbf{z}}
\newcommand{\govobj}{J}

\newcommand{\nonnegativenumbers}{\mathbb{R}_{\geq 0}}
\newcommand{\realnumbers}{\mathbb{R}}

\newcommand{\taxi}{\text{TX}} 
\newcommand{\pt}{\text{PT}} 
\newcommand{\costEnergy}{\phi^1} % cost_energy
\newcommand{\costLabour}{\phi^2} % cost_labour
\newcommand{\costVehicle}{\phi^3} % cost_license
\newcommand{\unitEmission}{\mu^1} % \unitEmission /km/pass

\newcommand{\tax}{\tau} % general tax/subsidy vector
\newcommand{\license}{\lambda}       % vehicle license
\newcommand{\subsidy}{\sigma} % general subsidy vector

\newcommand{\proj}{\operatorname{Proj}}
\newcommand{\argmin}{\arg\!\min}

\newcommand{\ite}{t} % subsidy for taxi

\definecolor{lightgreen}{RGB}{199, 237, 204}
\begin{document}
    \title{\textbf{Hierarchical Strategic Decision-Making in Layered Mobility Systems}}
    \author{
        Mingjia He$^{1,3}$, Zhiyu He$^{2}$, Jan Ghadamian$^{1}$, Florian Dörfler$^{2}$,  Emilio Frazzoli$^{1}$,  Gioele Zardini$^{3}$ %
        \thanks{This work was supported by the ETH Zürich Mobility Initiative (MI-03-22), the ETH Zürich Foundation (2022-HS-213), the Max Planck ETH Center for Learning Systems, and SNSF via NCCR Automation (51NF40\_225155).}%
        \thanks{$^{1}$Institute for Dynamic Systems and Control, ETH Zürich, 8092 Zürich, Switzerland 
        (e-mail: \texttt{\{minghe, ghjan,  efrazzoli\}@ethz.ch}).}%
        \thanks{$^{2}$Automatic Control Laboratory, ETH Zürich, 8092 Zürich, Switzerland 
        (e-mail: \texttt{\{zhiyhe, doerfler\}@control.ee.ethz.ch}).}%
        \thanks{$^{3}$Laboratory for Information and Decision Systems, Massachusetts Institute of Technology, Cambridge, MA, USA 
        (e-mail: \texttt{gzardini@mit.edu}).}%
    }
    % make the title area
    \maketitle

\thispagestyle{empty}
\pagestyle{empty}

%%%%%%%%%%%%%%%%%%%%%%%%%%%%%%%%%%%%%%%%%%%%%%%%%%%%%%%%%%%%%%%%%%%%%%%%%%%%%%%%
\begin{abstract}
Mobility systems are complex socio-technical environments influenced by multiple stakeholders with hierarchically interdependent decisions, rendering effective control and policy design inherently challenging. 
We bridge hierarchical game-theoretic modeling with online feedback optimization by casting urban mobility as a tri-level Stackelberg game (travelers, operators, municipality) closed in a feedback loop.
The municipality iteratively updates taxes, subsidies, and operational constraints using a projected two-point (gradient-free) scheme, while lower levels respond through equilibrium computations (Frank-Wolfe for traveler equilibrium; operator best responses).
This model-free pipeline enforces constraints, accommodates heterogeneous users and modes, and scales to higher-dimensional policy vectors without differentiating through equilibrium maps.

On a real multimodal network for Zurich, Switzerland, our method attains substantially better municipal objectives than Bayesian optimization and Genetic algorithms, and identifies integration incentives that increase multimodal usage while improving both operator objectives.
The results show that feedback-based regulation can steer competition toward cooperative outcomes and deliver tangible welfare gains in complex, data-rich mobility ecosystems.
% This paper bridges hierarchical game-theoretic modeling and online feedback optimization, demonstrating how policymakers can anticipate strategic responses while dynamically adjusting interventions to achieve social welfare objectives. We propose a hierarchical modeling framework to capture the strategic behaviors of users, mobility providers, and municipal authorities through interconnected decision layers. The overall hierarchical system with strategic interactions is viewed as a feedback loop, where the municipality leverages lower-level responses and iteratively refines its policies to optimize system-wide outcomes. To enable adaptive regulation, we employ a feedback optimization approach free of model information to identify effective policies targeting objectives, e.g., emission reduction and social welfare enhancement. We validate the proposed framework through a case study based on real-world travel demand and network data from Zurich, Switzerland. The results highlight the efficiency of our model-free approach in optimizing dynamic systems with complex interdependency and demonstrate how top-level interventions shape collaboration and competition among stakeholders.
\end{abstract}
%%%%%%%%%%%%%%%%%%%%%%%%%%%%%%%%%%%%%%%%%%%%%%%%%%%%%%%%%%%%%%%%%%%%%%%%%%%%%%%%

\section{Introduction}
Mobility systems are complex socio-technical infrastructures where users, operators, and policymakers interact strategically;
their collective choices determine congestion, emissions, costs, and welfare outcomes~\cite{martin2021av}.
%
%At the same time, as integral infrastructure, mobility systems are guided by broader societal goals that inform their development directions and specifications.
%For example, U.S.~and Chinese governments set concrete goals to improve transportation efficiency by reducing annual commuter delays, fleet emissions, and logistic costs
As public infrastructure, these systems are also steered by societal objectives codified in policy targets and regulations.
For instance, the U.S. and China have announced goals to reduce commuter delays, fleet emissions, and logistics costs~\cite{USDOT_NRSS_2022, China_TransportPlan_2022}.
To align private behavior with these objectives, policymakers deploy regulatory instruments such as taxes, subsidies, tolls, and service constraints~\cite{jalota2021efficiency}.

\begin{figure}[tb]
    \centering
    \includegraphics[width=0.6\linewidth]{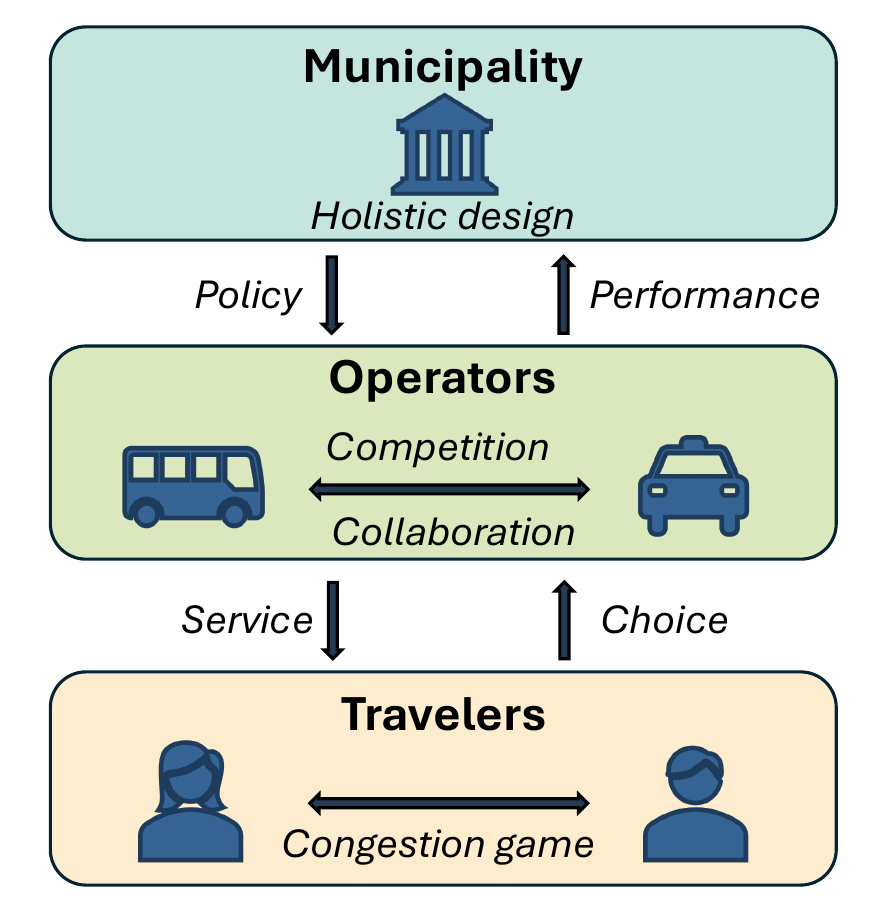}
    \caption{Hierarchical decision-making in mobility systems.}
    \vspace{-2ex}
    \label{fig:hierarchical_illustration}
\end{figure}

Effective control of mobility systems therefore hinges on understanding hierarchical strategic interactions across stakeholders (see \cref{fig:hierarchical_illustration}).
These interactions are both \emph{horizontal} and \emph{vertical}.
Horizontal interactions arise among agents at the same level.
On the user side, travel choices are interdependent through congestion externalities, a coupling captured by user-equilibrium traffic assignment models~\cite{kurmankhojayev2025methodological}.
%, liu2023end
On the operator side, platform decisions (e.g., pricing, capacity, service quality) may be \emph{cooperative}, for instance, integrating ride-pooling with bus networks~\cite{fielbaum2024improving} or deploying shared bikes for last-mile access~\cite{he2022geographically}, or \emph{competitive} due to pricing and platform differentiation~\cite{zhou2009pricing,kondor2022cost, zardini2022analysis,mo2021competition}.
Vertical interactions couple decisions across levels: operators shape demand-supply balance via pricing and resource allocation~\cite{schroder2020anomalous}, while municipalities influence the ecosystem through direct regulations and indirect economic incentives~\cite{gu2019tri,dandl2021regulating,zheng2023impacts}.

These structures are typically cast as bi-level Stackelberg programs with a single leader and multiple followers, aimed at designing optimal pricing policies that anticipate user-equilibrium responses~\cite{grontas2023,kim2025strategic}.
% It is common to model these structures using one-leader, multi-follower Stackelberg formulations or bi-level programs, often to design pricing policies that anticipate user-equilibrium responses~\cite{grontas2023,kim2025strategic}.
Other studies consider multi-leader, multi-follower settings to analyze interactions between multiple service providers and users, demonstrating how control interventions shape competition, cooperation, and welfare~\cite{gu2019tri,dandl2021regulating,zardini2021game,zardini2023strategic,bandiera2024mobility}. %salazar2018interaction,pantelidis2020many
%lanzetti2023interplay
Closely related to this paper,~\cite{zardini2021game,zardini2023strategic} develop general game-theoretic frameworks for layered urban mobility systems and analyze equilibrium outcomes under various policy scenarios, highlighting operator trade-offs between profitability and availability, and the resulting welfare implications.

Despite this progress, designing municipal policies that account for hierarchical coupling, strategic interactions, and cross-layer information flows remains challenging. 
First, lower-level responses (travelers and operators) often exhibit unknown parameters, non-smooth equilibrium mappings, and unmodeled couplings that undermine model-based approaches based on reformulation \cite{kim2025strategic} or explicit differentiation \cite{grontas2023}. 
Second, objectives are heterogeneous, and sometimes competing, across layers; 
upper-level decisions must anticipate heterogeneous lower-level reactions while enforcing constraints. Finally, sample-efficient hierarchical decision-making requires appropriate operating timescales of the municipality, mobility operators, and travelers.

In this work, we address these challenges through the emerging paradigm of feedback optimization \cite{simonetto2020time,hauswirth2021optimization}. At the heart of this paradigm is the closed-loop interconnection between optimization iterations and dynamical systems (e.g., mobility networks). This interconnection contributes to adaptation, performance optimization, and robustness against model mismatch and unforeseen disturbances. In our setting, we cast policy design as the interconnection between a decision maker (e.g., municipality) and lower-level strategic interactions (e.g., mobility operators and travelers). This decision maker leverages such strategic responses as feedback and employs gradient-free (two-point) updates with projections, thereby achieving online operations without lower-level modeling efforts \cite{Zhiyu2024}. Our approach naturally handles coupling constraints, scales to high-dimensional decision vectors, and remains robust to modeling error compared with heuristic exploration \cite{gu2019tri} and Bayesian optimization\cite{dandl2021regulating}.

\paragraph*{Contributions}
Our contributions are threefold.
% The contributions of this paper are threefold.
First, we propose a unified framework that captures both horizontal (traveler-traveler and operator-operator) and vertical (traveler-municipality-operator) strategic interactions.
Our modular framework offers sufficient granularity by incorporating heterogeneous classes of users and operators and their hierarchical interactions.  
% Building on prior game-theoretic formulations~\cite{zardini2021game,zardini2023strategic}, we incorporate hierarchical feedback loops between decision layers to obtain a controllable representation of mobility systems.
Second, we develop a model-free feedback-based regulatory design, in which municipalities iteratively refine taxes, subsidies, and operational constraints using gradient-free updates informed by observed lower-level responses.
Finally, we validate the proposed framework on real-world demand and network data from Zurich, Switzerland, showing improved efficiency over heuristic and Bayesian baselines and demonstrating how adaptive interventions approach optimality while shaping cooperation and competition among self-interested stakeholders.
%

% \zhmargin{My feedback on the introduction: i) shorten the second paragraph that discusses the importance of mobility system optimization (also remove some citations on URLs), ii) mention and compare with related work on bi/multi-level modeling and control of mobility systems, iii) summarize the challenges of hierarchical control in this setting (I can do that) and the algorithmic ideas.}

% \mhmargin{solved} 

\section{Mobility Interactions as a Stackelberg Game}
We model the interactions among the municipality, mobility operators, and travelers as a hierarchical decision-making process with three levels. 
At the top level, a municipality selects policy instruments that shape the overall mobility system. 
At the middle level, self-interested operators choose operational and pricing strategies in response to municipal policies and in anticipation of traveler behavior. 
At the low level, travelers make individual travel decisions based on the available services and their personal preferences.

% \zhmargin{minor: "low-level" since we do not have "lower v.s. upper"}
% \mhmargin{checked}

\subsection{Model Structure}
Let $\userset$ denote the set of travelers, and $\operatorset$ the set of mobility operators. 
The municipality chooses a control vector~$\govaction \in \govactionset$, where $\govactionset$ denotes the set of feasible policies. 
The strategy profiles for travelers and operators are denoted by 
$\useraction=\set{\useraction_\userindex}_{\userindex \in \userset} \in \useractionset$ 
and 
$\operatoraction= \set{\operatoraction_\operatorindex}_{\operatorindex \in \operatorset} \in \operatoractionset$, respectively.

% Each decision-maker has an associated cost function: $c_\userindex$ for traveler $\userindex$, $f_\operatorindex$ for TSP $\operatorindex$, and $\govobj$ for the government. 
% The dependence of each cost function on others' strategies introduces strategic coupling across levels. Throughout, the mappings $\mathrm{LNE}(\cdot)$ and $\mathrm{MNE}(\cdot)$ represent the lower- and middle-level equilibrium responses of travelers and TSPs, respectively, given upstream strategies.

\subsubsection{Low-Level Game}
Given the strategy profiles $\operatoraction$ of mobility operators and $\govaction$ of the municipality, travelers at the low level selfishly select routes and modes. Each traveler $\userindex \in \userset$ aims to minimize her own travel cost:
\begin{equation}
\min_{\useraction_\userindex \in \useractionset_\userindex}
\; \userobj_\userindex \tup{
\useraction_\userindex,
\useraction_{-\userindex}, \operatoraction,
\govaction},
\label{eq:lower}
\nonumber
\end{equation}
where $\userobj_\userindex: \useractionset \times \operatoractionset \times \govactionset \rightarrow \realnumbers$ denotes the cost function of traveler $\userindex$, 
which maps the joint strategy profile of all the travelers, operators, and the municipality into a nonnegative real value, and 
$\useraction_{-\userindex}$ denotes the strategy profile of all the travelers except $\userindex$.
This induces a noncooperative game among travelers. 
At equilibrium, no traveler can reduce their individual cost by unilaterally changing their strategy, given the strategies of others.

\begin{definition}[Low-Level Equilibrium]
\label{def:Lower-Level Equilibrium}
Given the strategy profile $\operatoraction$ of operators and the strategy $\govaction$ of the municipality, 
a strategy profile $\useraction^*$ is a \emph{Nash equilibrium} of the low-level traveler game if, for each traveler $\userindex \in \userset$,
\begin{equation}
c_\userindex \tup{
\useraction_\userindex^*,
\useraction_{-\userindex}^*,
\operatoraction,
\govaction
}
\;\leq\;
c_\userindex \tup{
\useraction_\userindex,
\useraction_{-\userindex}^*,
\operatoraction,
\govaction
},
\quad \forall\, \useraction_\userindex \in \useractionset_\userindex.
\label{eq:lne}
\end{equation}
\end{definition}
We denote by $\mathrm{LNE}: \operatoractionset \times \govactionset \rightarrow \useractionset$ the (possibly set-valued) mapping from the strategy profiles of the operators and the municipality 
to the corresponding equilibrium strategy profile of travelers, i.e.,
% \[
$\mathrm{LNE}(\operatoraction,\govaction)
:= \set{
    \useraction^* \in \useractionset
    \;\big|\;
    \useraction^* \;\text{satisfies}\; \eqref{eq:lne}
}.$
% \]

%
%
%

\subsubsection{Middle-Level Game}
Given the municipality policy $\govaction \in \govactionset$, mobility operators anticipate the travelers’ response and engage in a non-cooperative game. 
% \zhmargin{we should use municipality instead of government to keep consistency}
Each operator $\operatorindex \in \operatorset$ solves
\begin{equation}\label{eq:middle-level-operators}
\min_{\operatoraction_\operatorindex \in \operatoractionset_\operatorindex}
\quad 
\operatorobj_\operatorindex \tup{
    \operatoraction_\operatorindex,
    \operatoraction_{-\operatorindex},
    \useraction,
    \govaction
},
\end{equation}
where $\operatorobj_\operatorindex: \operatoractionset \times \useractionset \times \govactionset \rightarrow \realnumbers$ denotes the cost function of operator $\operatorindex$, and $\useraction \in \mathrm{LNE}\big((\operatoraction_\operatorindex^*,\operatoraction_{-\operatorindex}^*),\govaction\big)$ is the equilibrium response of travelers given the strategy profiles of mobility operators and the municipality.

\begin{definition}[Middle-Level Equilibrium]
Given $\govaction \in \govactionset$, 
a strategy profile $\operatoraction^*$ is a 
\emph{Stackelberg--Nash equilibrium} of the middle-level game if, 
for every operator $\operatorindex \in \operatorset$,
\begin{equation}
f_\operatorindex \big(
\operatoraction_\operatorindex^*,
\operatoraction_{-\operatorindex}^*,
\useraction^*,
\govaction
\big)
\!\leq\!
f_\operatorindex \big(
\tilde{\operatoraction}_\operatorindex,
\operatoraction_{-\operatorindex}^*,
\tilde{\useraction}^*,
\govaction
\big),
\quad
\forall\, \tilde{\operatoraction}_\operatorindex \in \operatoractionset_\operatorindex,
\label{eq:mne}
\end{equation}
where 
$\useraction^*\in \mathrm{LNE}(\operatoraction^*, \govaction)$ 
and
$\tilde{\useraction}^* \in \mathrm{LNE}
\tup{
    (\tilde{\operatoraction}_\operatorindex, \operatoraction_{-\operatorindex}^*), \govaction
}$.

\end{definition}

The mapping 
% \[
$\mathrm{MNE}(\govaction)
:=
\set{
    \operatoraction^* \in \operatoractionset
    \;\big|\;
    \operatoraction^*
    \text{ satisfies \eqref{eq:mne}}
}$
% \]
represents \emph{middle-level equilibrium}, indicating the equilibrium strategies of operators given the municipality policy $\govaction$.

\subsubsection{Top-Level Optimization}
Taking into account all interactions of operators and travelers in the low- and middle-level games, the municipality aims to guide the overall mobility system by choosing a control vector $\govaction \in \govactionset$ that minimizes an overall objective function:
\begin{equation}\label{eq:top_level_municipality}
\min_{\govaction \in \govactionset}
\; 
\govobj \tup{
    \govaction,
    \operatoraction,
    \useraction
}, \quad
\text{s.t. } \operatoraction \in \mathrm{MNE}(\govaction),~\useraction \in \mathrm{LNE}(\operatoraction, \govaction),
\end{equation}
where $\govobj : \govactionset \times \operatoractionset \times \useractionset \to \realnumbers$ is a function reflecting societal goals.
% \zhmargin{The top-level decision-making is optimization instead of game.}
We define the solution concept related to the above tri-level mobility system as follows.
\begin{definition}[Stackelberg Equilibrium of Mobility Game]\label{def:equi_tri_game}
A triplet 
$
(\govaction^*, \operatoraction^*, \useraction^*)
\in 
\govactionset 
\times 
\prod_{\operatorindex \in \operatorset} \operatoractionset_\operatorindex 
\times 
\prod_{\userindex \in \userset} \useractionset_\userindex
$
is a \emph{Stackelberg equilibrium} of the tri-level mobility system if the following conditions hold:
\begin{enumerate}
    \item
    The travelers’ strategies form a low-level equilibrium, i.e.,
    % \[
    $
    \useraction^*
    \in
    \mathrm{LNE}(\operatoraction^*, \govaction^*)$;
    % \]
    \item
    The operators’ strategies form a middle-level equilibrium given the municipality policy $\govaction^*$, i.e.,
    % \[
    $\operatoraction^*
    \in
    \mathrm{MNE}(\govaction^*)$;
    % \]
    \item
    The municipality strategy $\govaction^*$ minimizes its objective function, taking into account the induced middle- and low-level equilibria:
    % \begin{equation}
    $\govaction^*
    \in
    \argmin_{\govaction \in \govactionset}
    \govobj \tup{
    \govaction, \operatoraction^*, \useraction^*
    }$.
    % \nonumber
    % \label{eq:top}
    % \end{equation}
\end{enumerate}
\end{definition}

This tri-level Stackelberg formulation captures the hierarchical structure of decision-making in the mobility system, where municipality policies influence mobility operations, which in turn shape traveler behaviors. 
The resulting equilibrium
$(\govaction^*, \operatoraction^*, \useraction^*)$
represents a stable configuration in which no stakeholder has any incentive to unilaterally deviate from their chosen strategy.
% \zhmargin{I suggest merging Defs.~1-2 into Def.~3 to avoid repetition.}

\subsection{Mobility Game}
\subsubsection{Multimodal Mobility Network}
\label{sec:Multimodal Mobility Network}
We describe the mobility network as a directed labeled graph~$\graph=\tup{\nodes,\edges,\labelsmap}$,
where~$\nodes$ is the set of vertices,~$\edges \subseteq \nodes \times \nodes$ is the set of directed edges and~$\labelsmap: \edges \to  \labels $ is a mapping from the set of edges $\edges$ to the set of edge labels $\labels$.
%
%\zhmargin{$\labels$ is also used for government action set}
% change to O
We consider three transportation modes $\modeindex \in \modeset = \set{1,2,3}$, 
where $\modeindex=1,2$, and $3$ correspond to \gls{acr:pt}, road-based on-demand mobility services (e.g., taxis), and walking, respectively.
Private vehicle usage is treated as part of the default road flow.
To formalize multi-modality, the graph $\graph$ is structured as mode-specific subgraphs interconnected by mode-transfer edges. 
Specifically, for each transportation mode $\modeindex \in \modeset$, define a subgraph $\graph^\modeindex = (\nodes^\modeindex, \edges^\modeindex, \labelsmap^\modeindex)$.
The overall multimodal graph includes the union of these subgraphs and a set of intermodal transfer edges, i.e.,
$
\nodes = \bigcup_{\modeindex \in \modeset} \nodes^\modeindex
$ and
$
\edges = \left( \bigcup_{\modeindex \in \modeset} \edges^\modeindex \right) \cup \edges^0.
$
The set of mode-transfer edges $\edges^0$ enables transitions between modes and is defined as
$
\edges^0 \subseteq \bigcup_{\substack{\modeindex, \modeindex' \in \modeset}} 
\big( \nodes^{\modeindex} \times \nodes^{\modeindex'} \big),
$ with $\modeindex \neq \modeindex'$.

Each edge $e \in \edges$ carries a label $\labelsingle_e = (c_e, d_e, t_e) \in \labels = \nonnegativenumbers^3$.
%, 
% where $c_e$ denotes the monetary cost, 
% $d_e$ the travel distance, 
% and $t_e$ the travel or waiting time.
%
For a service edge $e \in \edges^k$ corresponding to a specific mode $\modeindex \in \modeset$, 
$c_e$ represents the mode-specific price, 
$d_e$ is the distance of the segment, 
and $t_e$ denotes the travel time under mode $k$. 
For a transfer edge $e \in \edges^0$,
% connecting nodes $v_i \in \nodes^{\modeindex_1}$ and $v_j \in \nodes^{\modeindex_2}$ with $k_1 \neq k_2$, 
$c_e$ captures the access or switching cost between modes, 
$t_e$ represents the expected transfer/waiting time,
and $d_e$ reflects the transfer distance. 
In the special case where the destination node lies in the walking layer, we set $c_e = 0$ and $t_e = 0$, meaning that transfers to walking are both cost-free and instantaneous. 
Additionally,  $d_e = 0$ if two vertices represent the same geographical location.

This layered representation captures mode-specific travel characteristics and inter- and intra-modal transfers at high granularity.
By explicitly including transfer edges (with associated access and waiting components) and mode-specific service edges, it supports precise path-cost evaluation that jointly accounts for time- and distance-based fares together with one-time access fees.

%%%
%
%
%

\subsubsection{Low-Level Game with Heterogeneous Travelers}
In the low-level game, we consider heterogeneous travelers, meaning that travelers differ in their preferences and perceptions of travel cost. 
Specifically, we adopt user group set $\usergroupset = \set{\text{Business}, \text{Commuting}, \text{Leisure}}$. Business travelers tend to prioritize travel time, leisure travelers prioritize monetary cost, while commuters have balanced preferences~\cite{Axhausen2006SwissVTT}.
Each group $\usergroupindex \in \usergroupset$ is characterized by a distinct value of time $\vot^\usergroupindex \in \nonnegativenumbers$, which reflects variations in income levels and sensitivity to time versus cost trade-offs.

%
%

%\zhmargin{No $\mathcal{I}$?}
Let $\request = (\usergroupindex,o,d,\requestnumber) \in \set{\usergroupset \times  \nodes \times \nodes \times \nonnegativenumbers} \triangleq \requestset$ denote trip requests,  
where $\usergroupindex$ denotes the user group, 
$o,d \in \nodes^3$ are the origin and destination nodes drawn from the walking layer,
and $\requestnumber$ is the number of requests.
A feasible path for a request $\request$ is defined as a subset of edges $\paths \subseteq \pathset_\request$ that forms a connected sequence from the origin vertex $o$ to the destination vertex $d$, where $\pathset_\request$ represents the set of all possible non-detour paths connecting $o$ to $d$.
The travel cost of a path $\paths$ consists of both monetary expenses and time valuation, and can be expressed as:
\begin{align*}
    C_{\paths}^{\usergroupindex} = 
    \sum_{e \in \paths} \tup{
    c_e + t_e \vot^\usergroupindex
    },
\end{align*}
where $c_e$ is the monetary cost on edge $e$, and $t_e$ is the associated travel time.
For \gls{acr:pt}, the travel time $t_e$ is a fixed value and independent of traffic.
For road-based services $e \in \edges^2$, the travel time $t_e$ is modeled by the Bureau of Public Roads (BPR) function~\cite{BPR}:
\begin{equation} 
\label{eq:BPR}
    t_e = t_e^0 \tup{1 + a  \tup{\frac{\flow_e+y_{\mathrm{def}}}{V_e}}^b}, \quad \forall e \in \edges^2, \nonumber
\end{equation}
%pt waiting time = 60/fre
%TX_waiting_time = -29 / 900 * FWvar["TX"]["w"].get_value() + 299 / 9
where $t_e^0$ is the free-flow-time that it takes to travel on link $e$ without any form of congestion; 
$a$ and $b$ are parameters related to the BPR function (the standard values $a=0.15$ and $b=4$ were chosen \cite{BPR}); $\flow_e$ describes the traveler flow on edge $e$, and $y_{\mathrm{def}}$ is private car flow; $V_e$ describes the capacity on link $e$.
% \zhmargin{It would be great to discuss the relationship between $\flow$ and $\useraction$}
%
Each traveler $\userindex$ selects a path $\useraction_\userindex$ that minimizes their perceived travel cost $C_{\paths}^{\usergroupindex}$, i.e., $\useraction_\userindex \in \arg\min_{\paths \in \mathcal{P}_\request} C_{\paths}^{\usergroupindex}$. 
% Formally, the traveler strategy $\useraction_\userindex$ is the chosen path,
% \begin{align}
%     \useraction_\userindex \in \arg\min_{\paths \in \mathcal{P}_\request} C_{\paths}^{\usergroupindex}. \nonumber
% \end{align}
The collection of individual path choices $(\useraction_\userindex)_{\userindex\in\userset}$ induces an aggregate link-flow vector
$\flow=(y_e)_{e\in\edges}\in\nonnegativenumbers^{|\edges|}$ in the low-level model. The traveler flow on edge $e$ satisfies $y_e = \sum_{\userindex\in\userset} \delta_e^{\useraction_\userindex}$,
% \begin{align}
%     y_e = \sum_{\userindex\in\userset} \delta_e^{\useraction_\userindex},\nonumber
% \end{align}
% That is, the flow on each link $e$ equals the number of travelers whose chosen path uses link $e$. Hence, $\flow$ is the aggregate outcome generated by the individual equilibrium decisions $\useraction$.
where $\delta_{e}^{\useraction_\userindex}$ is a binary indicator equal to $1$ if edge $e$ is contained in the chosen path $\useraction_\userindex$, and $0$ otherwise.
According to \cref{def:Lower-Level Equilibrium}, this equilibrium flow can be formulated as a variational inequality~\cite{dafermos1980traffic} and equivalently obtained from the following optimization problem:
\begin{subequations}\label{prob:LL}
\begin{align}
\min_{\flow} \quad & \sum_{e \in \edges} \int_{0}^{\flow_e} t_e(w) \, dw 
+ \sum_{e \in \edges} \sum_{\usergroupindex \in \usergroupset} \frac{\flow^\usergroupindex_e}{\vot^\usergroupindex} c_e, \label{eq:LL_obj} 
\\
\text{s.t.} \quad & \sum_{\paths \in \pathset_\request} f^\request_\paths = q_\request, \quad \forall \request \in \requestset, \label{eq:LL_flow_conservation}
\\
&  \sum_{\substack{\request \in \requestset \\ \usergroupindex_\request = \usergroupindex}}
    \sum_{p \in \pathset_\request}
    \delta_{e}^p f_p^\request = \flow^\usergroupindex_e, 
    \quad \forall e \in \edges, \; \forall \usergroupindex \in \usergroupset,
\label{eq:LL_edge_flow_by_group}
\\
& \sum_{\usergroupindex \in \usergroupset} \flow^\usergroupindex_e = \flow_e, \quad \forall e \in \edges.
\label{eq:LL_edge_total_flow}
\end{align}
\end{subequations}
% update the previous section about y
%
In the objective function \eqref{eq:LL_obj}, for the multi-class traveler setting, we retain the Beckmann potential term and convert the monetary costs into time units using the value of time for each user group. Furthermore, $f^\request_\paths \in \nonnegativenumbers$ represents the flow of request $r$ choosing a path $p\in\mathcal P_r$;
$\requestnumber_\request$ denotes the number of requests $r$;
Constraint \eqref{eq:LL_flow_conservation} enforces flow conservation, ensuring that the demand associated with each request is fully allocated across feasible paths.
Constraints \eqref{eq:LL_edge_flow_by_group} and \eqref{eq:LL_edge_total_flow} define the edge flows for each traveler group and the total edge flows aggregated across all groups, respectively.

% The number of licensed vehicles affects the vehicle turnover rate, thereby influencing average passenger waiting times. 
% Furthermore, the taxi operator is subject to capacity constraints from the number of licensed vehicles in operation (???).

%\mh{Analyze: the choice of congestion function}\\
%\mh{Analyze: the existence and uniqueness of the congestion game}
%\theorem{Existence and uniqueness of the congestion game}\\

\subsubsection{Middle-Level Game for Service Operators}

We consider two transportation modes: the \gls{acr:pt} and \gls{acr:tx} service, denoted by $\operatorindex \in\operatorset \triangleq \{\text{\gls{acr:pt}}, \text{\gls{acr:tx}}\}$.

\noindent
\textit{Strategy:}
The joint strategy profile of operators can be represented by a vector 
$\mathbf{\operatoraction} = \tup{ \mathbf{\operatoraction_{\pt}},\, \mathbf{\operatoraction_{\taxi}}} \in \operatoractionset $.
The strategy of the public transport operator is given by
$\mathbf{x}_{\pt} = \left(\ptfre,\; p_{\pt}^{\text{base}},\;p_{\pt}^{d},\; p_{\pt}^{\text{trans}}\right) \in \mathbb{R}_{\geq 0}^4$, 
where the public transport operator determines the service frequency $q$, 
the base fare $p_{\pt}^{\text{base}}$, 
the distance-based price $p_{\pt}^{d}$, 
and the transfer cost to users switching from taxi services $p_{\pt}^{\text{trans}}$.
The strategy of the taxi operator is
$\mathbf{x}_{\taxi} = \left( w,\; p_{\taxi}^{\text{base}},\; p_{\taxi}^d,\; p_{\taxi}^t,\; p_{\taxi}^{\text{trans}}\right) 
\in \mathbb{R}_{\geq 0}^5$,
where the taxi operator determines the number of operated vehicles $w$, 
base fare $p_{\taxi}^{\text{base}}$, 
distance-based pricing $p_{\taxi}^d$,
time-based price $p_{\taxi}^t$ and 
the transfer cost for passengers transferring from public transport $p_{\taxi}^{\text{trans}}$. 
It should be noted that both operators are allowed to adjust the transfer price, which reflects their intention to integrate with the other mobility service. For this consideration, this price will not be larger than the base fare of the corresponding service.
% \zhmargin{We should highlight that transfer costs are related to collaboration, i.e., in practice they are lower than base prices.}
% checked

% All the pricing variables $ \mathbf{p}$ are subject to upper bound constraints $\mathbf{p}_{\max}$.
% Additionally, the service frequency for public transport is constrained by the maximum frequency $\ptfre_{\max}$, 
% and the number of operated vehicles for taxi operators is limited by the number of issued vehicle licenses.
% \begin{align}
% \left(\mathbf{p}, q, w \right) \leq \left(\mathbf{p}_{\max}, q_{\max}, \license \right) \label{eq:operator_constraints}
% \end{align}

%
%
%
\noindent
\textit{Performance Metrics:}
We evaluate operator performance using two key metrics: revenue and operational cost. 
The total revenue of operator $\operatorindex \in \{\text{\gls{acr:pt}}, \text{\gls{acr:tx}}\}$ is given by:
\begin{align}
    \operatorobj^\mathrm{rev}_\pt (\operatoraction,\useraction)  & =
    \sum_{e \in \edges^1} \flow_e l_e p_{\pt}^d
            +\sum_{e \in \edges^{0}_{31}} \flow_e  p_{\pt}^\mathrm{base}
            +\sum_{e \in \edges^{0}_{21}} \flow_e  p_{\pt}^\mathrm{trans}, \nonumber\\
    \operatorobj^\mathrm{rev}_\taxi (\operatoraction,\useraction)  & =
    \sum_{e \in \edges^2} \flow_e \tup{l_e p_{\taxi}^d+t_e p_{\taxi}^t} \nonumber\\
            &\quad + \sum_{e \in \edges^{0}_{32}} \flow_e  p_{\taxi}^\mathrm{base}
            +\sum_{e \in \edges^{0}_{12}} \flow_e  p_{\taxi}^\mathrm{trans} \nonumber,
\end{align}
where $\flow_e$ is the number of users on edge $e$. 
The total cost for each operator is modeled as the sum of distance-related and time-related costs over all relevant edges, i.e.,
\begin{align}
    \operatorobj^\mathrm{cost}_{\pt} (\operatoraction) & = 
    \ptfre \sum_{e \in \edges^1} \tup{l_e \costEnergy_{\pt} + t_e\costLabour_{\pt} }
    + \costVehicle_{\pt} \ptfre , \nonumber\\
    \operatorobj^\mathrm{cost}_{\taxi} (\operatoraction, \useraction) & =  
    \sum_{e \in \edges^2} \flow_e \tup{l_e \costEnergy_{\taxi} + t_e\costLabour_{\taxi}}
    + \costVehicle_{\taxi} w, \nonumber
\end{align}
% where $\costEnergy$ and $\costLabour$ denote the distance-related unit cost (e.g., energy) and time-related unit cost (e.g., labor), respectively. For \gls{acr:pt}, the cost is scaled by the service frequency $\ptfre$, while for \gls{acr:tx}, the cost depends on travel demand $\useraction$, as taxis operate only where passengers request service. This formulation can also be extended to incorporate emission costs or other operational considerations.
where $\costEnergy$ and $\costLabour$ denote distance- and time-related unit costs (e.g., energy and labor). For \gls{acr:pt}, the cost scales with service frequency $\ptfre$, while for \gls{acr:tx}, it depends on travel demand $\useraction$, since taxis operate only where requested. This formulation can also be extended to incorporate emission costs or other operational considerations.

\noindent
\textit{Self-Interested Operators:}
Each operator solves its own optimization problem given the municipality strategy $\govaction$, and the travelers’ equilibrium response. 
Municipal interventions influence operator objectives through three primary mechanisms.
(i) $\tax = (\tax_{\pt}, \tax_{\taxi})$ denotes the revenue-related policy, with positive values representing tax ratios and negative values representing subsidy ratios applied to operators;
(ii) the licensing variable $\license$ regulates the number of operational taxi vehicles.
and (iii) the subsidy component $\subsidy = (\subsidy_{\pt}, \subsidy_{\taxi})$ provides per-passenger incentives to promote mobility service integration and encourage shifts toward specific transport modes.
The optimization problem for \gls{acr:pt} operator can be expressed as:
\begin{subequations}\label{eq:ML_public_prob}
    \begin{align}
    \min_{\operatoraction_\pt}  \quad & \operatorobj_\pt  = (\tax_\pt\!-\!1) \operatorobj^\mathrm{rev}_\pt + \operatorobj^\mathrm{cost}_\pt - \subsidy_\pt \sum_{e \in \edges^0_{21}} \flow_e \label{eq:operator_obj_pt}\\
    \text{s.t.} \quad & \useraction \in \mathrm{LNE}(\operatoraction, \govaction), \\
    & \left(\ptfre,\; p_{\pt}^{\text{base}},\;p_{\pt}^{d},\; p_{\pt}^{\text{trans}}\right) \leq 
    \mathbf{P}_{\pt}. \label{eq:operator_bound}
\end{align}
\end{subequations}
% \zhmargin{highlight competing decisions}
Similarly, the optimization problem for \gls{acr:tx} operator is:
\begin{subequations}\label{eq:ML_taxi_prob}
\begin{align}
    \min_{\operatoraction_\taxi}  \quad & \operatorobj_\taxi  = (\tax_\taxi\!-\!1) \operatorobj^\mathrm{rev}_\taxi + \operatorobj^\mathrm{cost}_\taxi - \subsidy_\taxi \sum_{e \in \edges^0_{12}} \flow_e \label{eq:operator_obj_tx}\\
    \text{s.t.} \quad & \useraction \in \mathrm{LNE}(\operatoraction, \govaction), \\
    & \left( p_{\taxi}^{\text{base}}, p_{\taxi}^d, p_{\taxi}^t, p_{\taxi}^{\text{trans}}\right)  \leq 
     \mathbf{P}_{\taxi}, \label{eq:operator_taxi_bound} \\
     & w  \leq \license. \label{eq:ML_license}
\end{align}
\end{subequations}
In \eqref{eq:operator_obj_pt}, the multiplication of the revenue $\operatorobj^\mathrm{rev}_\pt$ and the coefficient $\tax_\pt\!-\!1$ gives the negative revenue after taxation. In \eqref{eq:operator_bound} and \eqref{eq:operator_taxi_bound}, $\mathbf{P}_{\pt}$ and $\mathbf{P}_{\taxi}$ are the upper bounds on the decision variables. Further, in \eqref{eq:ML_license}, the number of operating taxis $w$ is constrained by the license bound regulated by the municipality, see the following specification. Further, let $\operatoractionset_{\pt} \triangleq \{\operatoraction_\pt \in \times \mathbb{R}_{\geq 0}^3 | \operatoraction_\pt \text{ satisfies } \eqref{eq:operator_bound}\}$ and $\operatoractionset_{\taxi} \triangleq \{\operatoraction_\taxi \in \mathbb{R}_{\geq 0}^5 | \operatoraction_\taxi \text{ satisfies } \eqref{eq:operator_taxi_bound}, \eqref{eq:ML_license}\}$ denote the constraint sets for $\operatoraction_\pt$ and $\operatoraction_\taxi$, respectively. The strategic interactions between self-interested \gls{acr:pt} and \gls{acr:tx} operators characterized by problems \eqref{eq:ML_public_prob} and \eqref{eq:ML_taxi_prob} constitute special cases of the middle-level game \eqref{eq:middle-level-operators}.

% \zhmargin{We should point out that $\tau$ represents the percentage, and $(\tau-1)*rev$ gives the negative actual revenue after taxation.}

\subsubsection{Top-Level Optimization for Municipality}
The municipality seeks to enhance the overall system performance by improving social welfare, 
reducing emissions, and minimizing public expenditure. 
To achieve these goals, the municipality searches for an optimal control vector
$\govaction=\tup{\tax,\license,\subsidy} \in \mathbb{R}^2 \times \mathbb{R}_{\geq 0} \times \mathbb{R}^2_{\geq 0}$, 
whose elements represent revenue-related policy, vehicle licensing, and subsidies, respectively.

The optimization problem for the municipality is represented by:
% \begin{subequations}\label{eq:municipality_prob}
%     \begin{align}
%     \min_{\govaction} \quad 
%     \govobj(\govaction) &= - \omega_1 \govobj^\mathrm{sw}
%     + \omega_2 \govobj^\mathrm{em}
%     - \omega_3  \govobj^\mathrm{rev} \label{eq:social_obj} \\
%     \text{s.t.} \quad 
%     & (\operatoraction, \useraction) \in \mathrm{MNE}(\govaction), \\
%     & \tau_\operatorindex \in [\tau^\operatorindex_{\min}, \tau^\operatorindex_{\max}],  \label{eq:tax_bound} \\
%     & \lambda \in [0, \lambda_{\max}], \label{eq:lambda_bound} \\
%     & \subsidy_\operatorindex \in [0, \subsidy^\operatorindex_{\max}], \label{eq:subsidy_bound}
%     \end{align}
% \end{subequations}
\begin{subequations}\label{eq:municipality_prob}
\begin{align}
\min_{\govaction}\quad 
& \govobj(\govaction)= -\omega_1 \govobj^\mathrm{sw} + \omega_2 \govobj^\mathrm{em} - \omega_3 \govobj^\mathrm{rev} \label{eq:social_obj}\\
\text{s.t.}\quad 
& (\operatoraction,\useraction)\in \mathrm{MNE}(\govaction), \\
& \govaction \in [\tau_{\min},\tau_{\max}]\times[0,\lambda_{\max}]\times[0,\subsidy_{\max}], \label{eq:constr_z_municipality}
\end{align}
\end{subequations}
where $\govobj^\mathrm{sw}$ denotes social welfare, defined as the total travel cost of the system in \eqref{eq:TL_municipality_sw}; $\govobj^\mathrm{em}$ represents the emissions generated by taxi services, calculated using \eqref{eq:TL_emission_taxi}, with $\unitEmission$ denoting the unit emission rate of the taxi service; and $\govobj^\mathrm{rev}$ denotes the revenue of the municipality, given by the difference between the taxes from the operators and the subsidy and specified in \eqref{eq:TL_municipality_rev}:
\begin{subequations}
\begin{align}
    & \govobj^\mathrm{sw} (\govaction) =  \sum_{\usergroupindex \in \usergroupset} \sum_{e \in \edges} \flow_e^\usergroupindex \tup{c_e + t_e \vot^\usergroupindex},
    \label{eq:TL_municipality_sw} \\
    & \govobj^\mathrm{em}(\govaction) = \sum_{e \in \edges^{2}} \flow_e l_e \unitEmission, 
    \label{eq:TL_emission_taxi}  \\
    & \govobj^\mathrm{rev}(\govaction) = \sum_{\operatorindex \in \operatorset} \tax_\operatorindex \operatorobj^\mathrm{rev}_\operatorindex 
    - \subsidy_\operatorindex \sum_{e \in \edges^{0}_{m' m^k}} \flow_e.
    \label{eq:TL_municipality_rev}  
\end{align}
\end{subequations}
Moreover, $\omega_1,\omega_2,\omega_3 \in \mathbb{R}_{\geq 0}$ in \eqref{eq:social_obj} are weighting parameters. Let $\govactionset \triangleq \{\govaction \in \mathbb{R}^2 \times \mathbb{R}_{\geq 0} \times \mathbb{R}^2_{\geq 0} | \govaction \text{ satisfies } \eqref{eq:constr_z_municipality}\}$ denote the constraint set. Then, problem \eqref{eq:municipality_prob} corresponds to the general formulation \eqref{eq:top_level_municipality}.

% \zhmargin{check objectives}

% !TEX root = ..\root.tex
\section{Solving the Hierarchical Mobility Game}
% \section{Solving Mobility Games: A Model-Free Feedback-based Approach}

The hierarchical mobility game couples three layers of strategic decision making. 
Computing a Stackelberg equilibrium directly is difficult due to nested optimization, strategic coupling, and limited model knowledge.
We therefore adopt a \emph{model-free, feedback-based} approach: each layer updates its decision by probing the system with small, structured perturbations and using observed performance to adjust in closed loop.
% for hierarchical mobility systems.

\subsection{Challenges of Hierarchical Decision-Making}
The difficulties in computing the Stackelberg equilibrium of the tri-level system as per \cref{def:equi_tri_game} are threefold.

\begin{itemize}
    \item \textbf{Implicit lower-level responses.} The mappings $\mathrm{LNE}(\operatoraction,\govaction)$ and $\mathrm{MNE}(\govaction)$ are generally unavailable in closed form and are cumbersome to characterize via KKT conditions or variation inequalities, which limits direct differentiation or exact reformulations.
    \item \textbf{Nonsmooth, nested objectives.} The municipality’s and operators’ objectives depend on equilibrium selections of downstream layers. 
    This induces nonsmooth, nonconvex, and nested dependencies that challenge standard model-based pipelines relying on explicit gradients through layers~\cite{blondel2024elements}.
    \item \textbf{Constraints and uncertainty.} Decisions are constrained (budgets, prices, capacities), noises and modeling errors perturb measured outcomes, and inner equilibria may be computed only approximately.
\end{itemize}
% \begin{itemize}
%     \item First, the low-level and middle-level equilibrium mappings lack closed-form expressions, which inhibits the direct use of numerical solvers. Also, these mappings are hard to characterize through optimality conditions, complicating reformulation via equilibrium constraints.

%     \item Second, the overall hierarchical system involves complex interactions represented by nested optimization, interdependent variables, and non-smooth mappings (related to $\mathrm{LNE}(\operatoraction, \govaction)$ and $\mathrm{MNE}(\govaction)$). Such interactions are unfavorable for typical \emph{model-based} pipelines that rely on precise expressions and parameters to differentiate through layers \cite{blondel2024elements} and perform iterative updates.
%     \zhmargin{refs}

%     % \item Finally, the decision timescales of stakeholders at different levels should be treated with care. To facilitate adjustments and adaptation, for each decision vector $\govaction$, the top-level municipality requires approximate knowledge of operator strategies $\mathbf{\operatoraction}$ and travel flows $\useraction$ at equilibrium. Similarly, given municipality specifications, self-interested operators need (approximate) equilibrium flows to iteratively optimize individual objectives. Choosing appropriate updating timescales for different stakeholders is crucial for convergence of adjustments within layers.
% \end{itemize}

To address the above challenges, we present a hierarchical decision-making pipeline based on model-free feedback optimization~\cite{Zhiyu2024}. 
The key idea is to exploit responses (i.e., decision variables when close to equilibria) of lower-level stakeholders as feedback and perform iterative adjustments based on structured exploration and objective evaluations.

% The presented hierarchical mobility game features intra/inter-level strategic interactions and brings unique challenges to the overall decision-making. The interactions between mobility service providers and travelers are represented by multi-leader-multi-follower games. The transformatidon of these games as equilibrium constraints is non-trivial. Moreover, there lack closed-form expressions for the mappings of lower-level and middle-level Nash equilibria. This lack of information inhibits directly using numerical solvers or gradient schemes based on automatic differentiation (requiring model parameters). Finally, the objective functions involve non-smooth mappings and bilinear terms, thus complicating iterative updates and asymptotic convergence.

\subsection{Solving Low-Level Equilibria}
Given operator actions~$\operatoraction$ and municipal policy~$\govaction$, the equilibrium flow~$\flow \in \mathrm{LNE}(\operatoraction,\govaction)$ is computed by solving the convex Beckmann reformulation of \cref{prob:LL} and applying the Frank--Wolfe (FW) algorithm~\cite[Ch~6.2]{Boyles2022}.
We allow heterogeneous traveler classes~$\usergroupindex\in\usergroupset$ (e.g., values of time, access fees), with class-specific demands and generalized link costs.
As summarized in \cref{alg:fw}, 
at each iteration $\ite$, \texttt{SPAssignment} computes an auxiliary flow~$\flow^{*,\ite}$ by solving a shortest-path problem for each origin--destination pair and each traveler class. This shortest-path problem is concerned with minimizing the generalized travel cost and is solved via Dijkstra’s algorithm. Next, the optimal step size $\alpha^\ite \in [0,1]$ is determined through a line-search procedure \texttt{find$\alpha$}. The flow is then updated as
$\flow^{\ite+1} = (1-\alpha^\ite)\flow^\ite + \alpha^\ite \flow^{*,\ite}$, see \texttt{updateFlow\&Cost}.
\texttt{getTSTC} computes the total system travel time associated with the updated flow.
This process continues until the convergence criterion (e.g., $\frac{t^{\text{ts},\ite}}{t^{\text{sp},\ite}} - 1 < \varepsilon$) is satisfied, 
at which point the total system travel time $t^{\text{ts},\ite}$ and the shortest-path travel time $t^{\text{sp},\ite}$ become sufficiently close, indicating that the user equilibrium is reached.

% \zhmargin{Should the returned variable be the flow?}
% checked

% \zhmargin{change $k$ because it is used for the operator index}
% use $s$ 

% The algorithmic updates are:
% \begin{enumerate}
%     \item Generate a target solution $\flow^{*t}$ via a shortest-path (e.g., Dijkstra) algorithm; \label{step:generate_sol}
%     \item Determine $\alpha^t$ based on \ldots
%     \item Update the current flow by $\flow^{t+1} = \alpha^t \flow^{*t} + (1-\alpha^t) \flow^t$;
%     \item Check the stopping criterion and goes back to step 1) if necessary.
% \end{enumerate}

% \zh{please add more details}\\
% \mh{MH: updated, please check}

\begin{algorithm}[H]
\caption{Frank--Wolfe for the low-level congestion game}
\label{alg:fw}
\small
\begin{algorithmic}[1]
\Require Network $\graph^0$, operators' action $\operatoraction$, demand $\requestset$
\State $t^{\text{ts},0} \leftarrow +\infty$, $(t^{\text{sp},0}, y^{*,0}) \leftarrow \texttt{SPAssignment($\graph^0$)}$
\While{$\frac{t^{\text{ts},\ite}}{t^{\text{sp},\ite}}-1 \ge \varepsilon$}
    \State $\alpha^\ite \leftarrow \texttt{find$\alpha$}(\graph^t, y^{*,\ite})$
    \State $(y^\ite,\graph^\ite) \leftarrow \texttt{updateFlow\&Cost}(\graph^t, y, x,\alpha^\ite)$
    \State $(t^{\text{sp},\ite}, y^{*,\ite} )\leftarrow \texttt{SPAssignment($\graph^\ite$})$
    \State $t^{\text{ts},\ite} \leftarrow \texttt{getTSTC($\graph^\ite$)}$
    \EndWhile
\State \Return $y^\ite, t^{\text{ts},\ite}$
\end{algorithmic}
\vspace{0.5ex}
\footnotesize
\end{algorithm}

\subsection{Model-Free Iterations for Middle and Top Levels}
We present the iterative update rules adopted by operators and the municipality to compute the middle-level equilibrium and the top-level optima in the hierarchical mobility game, respectively. These rules are \emph{model-free}, in that (approximate) equilibrium responses from lower levels rather than their explicit mappings or accurate parameters are employed. Furthermore, the concise structure of such rules facilitates numerical implementation and constraint satisfaction.  

% To address the above challenges, we leverage a model-free feedback optimization approach. Specifically, the decision maker at each level views the interactions below the current layer as a block box. Then, iterative updates are performed by interacting with lower levels through structured exploration and descent-based updates.

\subsubsection{Middle-Level Game}
Given the top-level municipality policy $\govaction$, operators seek the middle-level equilibrium by repeatedly updating strategies based on the corresponding low-level equilibrium flows. Specifically, the iterative rule adopted by the middle-level operator $\operatorindex \in \{\text{\gls{acr:pt}}, \text{\gls{acr:tx}}\}$ is
\begin{subequations}\label{eq:middle_operator_itr}
\begin{align}
    \operatoraction_\operatorindex^{t+1} &\!=\! \proj_{\operatoractionset_{k}}(\operatoraction_\operatorindex^t - \eta g_\operatorindex^t), \label{eq:middle_operator_update} \\
    g_\operatorindex^t &\!=\! \frac{v^t}{2\delta} \left(\widehat\operatorobj_\operatorindex(\operatoraction_\operatorindex^t \!+\! \delta v^t; \operatoraction_{-\operatorindex}^t) \!\!-\!\! \widehat\operatorobj_\operatorindex(\operatoraction_\operatorindex^t \!-\! \delta v^t; \operatoraction_{-\operatorindex}^t) \right). \label{eq:middle_operator_exploration}
\end{align}
\end{subequations}
In \eqref{eq:middle_operator_update}, $t \in \mathbb{N}$ is the iteration count, $\operatoractionset_\operatorindex$ is the constraint set (see the discussions after \eqref{eq:ML_taxi_prob}), $\proj_{\operatoractionset_\operatorindex}(\cdot)$ denotes the projection onto $\operatoractionset_\operatorindex$, and $\eta > 0$ is a constant step size. In \eqref{eq:middle_operator_exploration},  $v^t \sim \mathcal{N}(0,I)$ is a standard normal Gaussian vector for exploration, $\delta$ is a parameter that controls the exploration range, and $g_\operatorindex^t$ is an updating direction obtained by scaling $v^t$ with the difference of two objective evaluations $\widehat\operatorobj_\operatorindex$. 

% Specifically, given the municipality policy $\govaction$, the opponent strategy $\operatoraction_{-\operatorindex}^t$, and the perturbed strategy $\operatoraction_\operatorindex^t \! \pm \! \delta v^t$, we use \cref{alg:fw} to obtain an (approximate) equilibrium flow and then evaluate as per \eqref{eq:operator_obj} or \eqref{eq:operator_obj_tx}.

We explain the implementation and intuition of \eqref{eq:middle_operator_itr} as follows. At each iteration $t$, \gls{acr:pt} and \gls{acr:tx} operators adjust their decisions $\operatoraction_\pt$ and $\operatoraction_\taxi$ while respecting the constraint sets $\operatoractionset_{\pt}$ and $\operatoractionset_{\taxi}$ through projection. The updating direction $g_\operatorindex^t$ approximates the partial gradient $\nabla_{\operatoraction_\operatorindex} \operatorobj_\operatorindex(\operatoraction_\operatorindex,\operatoraction_{-\operatorindex})$ (appearing in gradient play \cite{BasarOlsder1998}), thus facilitating equilibrium seeking. To construct $g_\operatorindex^t$, the operator $\operatorindex$ explores her own objective $\operatorobj_\operatorindex$ (see \eqref{eq:operator_obj_pt} or \eqref{eq:operator_obj_tx}) around the current decision $\operatoraction_\operatorindex^t$ and scales $v^t$ with the difference of objective evaluations, a two-point form typical in zeroth-order optimization \cite{liu2020primer}. Evaluation of $\widehat\operatorobj_\operatorindex$ requires solving for the (approximate) low-level equilibrium flow via \cref{alg:fw} given municipality policy $\govaction$, the opponent strategy $\operatoraction_{-\operatorindex}^t$ and the perturbed strategy $\operatoraction_\operatorindex^t \! \pm \! \delta v^t$.
% use \cref{alg:fw} to obtain an (approximate) equilibrium flow and then evaluate as per \eqref{eq:operator_obj} or \eqref{eq:operator_obj_tx}

The iterative rule \eqref{eq:middle_operator_itr} features \emph{model-free} operations, in that middle-level operators do not access the detailed parameters (e.g., the number of requests $\requestnumber_\request$ and the value of time $\vot^\usergroupindex$) or the structure of the low-level congestion game \eqref{prob:LL}. Instead, only (approximate) equilibrium flows are exploited to construct two-point gradient estimates as per \eqref{eq:middle_operator_exploration}. The scheme \eqref{eq:middle_operator_itr} is also simple to implement via projections onto box constraints and other standard operations.

% Although evaluating $\operatorobj_\operatorindex$ based on the low-level equilibrium flow can be intensive, we will demonstrate in \cref{sec:numerical} that the middle-level update requires moderate iterations to reach the equilibrium $\mathrm{MNE}(\govaction)$ in a real-world large-scale setting.

% \zh{middle-level approximation using neural networks}

% discuss in \cref{subsec:alg_extension} the tailored stopping strategy to reduce the computational load

\subsubsection{Top-Level Optimization}
In a similar vein, the top-level municipality employs an iterative rule based on gradient estimates and (approximate) equilibrium responses of the middle- and low-level stakeholders. The update rule is
\begin{subequations}\label{eq:top_operator_itr}
\begin{align}
    \govaction^{t+1} &= \proj_{\govactionset}(\govaction^t - \eta g_\govaction), \\
    g_\govaction^{t} &= \frac{v^t}{2\delta}\left(\widehat\govobj(\govaction^t \!+\!\delta v^t) - \widehat\govobj(\govaction^t \!-\!\delta v^t) \right), \label{eq:top_level_grad_est}
\end{align}
\end{subequations}
where $\govactionset$ is the municipality action set, $\proj_{\govactionset}(\cdot)$ denotes the projection onto $\govactionset$, and the step size $\eta$, the scaling parameter $\delta$, and the exploration noise $v^t \sim \mathcal{N}(0,I)$ retain the same interpretations as those in \eqref{eq:middle_operator_itr}. Moreover, $\widehat\govobj(\govaction^t  \pm \delta v^t)$ denotes an evaluation of the municipality objective $\govobj$ at the perturbed policy vector $\govaction^t \pm \delta v^t$.
% the objective $\govobj$ is given in \eqref{eq:social_obj}

The municipality iteratively adjusts the policy $\govaction$ while ensuring constraint satisfaction $\govaction \in \govactionset$. To construct a gradient estimate \eqref{eq:top_level_grad_est}, the municipality bypasses middle- and low-level problem parameters (e.g., cost factors and travel requests) and structures. Instead, the municipality leverages (approximate) evaluations of the overall objective $\govobj$ based on the middle- and low-level responses. Theoretically speaking, given the municipality policy $\govaction^t \pm \delta v^t$, we may call multiple rounds of \eqref{eq:middle_operator_itr} for middle-level operators to approach $\mathrm{MNE}(\govaction)$ and further obtain $\widehat\govobj(\govaction^t \pm \delta v^t)$. In numerical evaluations in \cref{sec:numerical}, we leverage a pre-trained neural-network approximation of $\mathrm{MNE}(\govaction)$ to save online evaluation effort. Specifically, we sample sufficiently many combinations of the municipality policy $\govaction$ and the operator strategies and solve for equilibrium responses via \eqref{eq:middle_operator_update}. Building on such data constructed offline, we fit a multi-layer perceptron (MLP) regressor approximating $\mathrm{MNE}(\govaction)$. During online inference, this learned regressor generates (approximate) equilibrium responses, which are then employed to evaluate $\widehat\govobj(\govaction^t  \pm \delta v^t)$.

We summarize the above model-free iterative rules in \cref{alg:middle-equilibrium,alg:top-opt}. 
In particular, \texttt{evaluate$\operatorobj_\operatorindex$} and \texttt{evaluate$J$} compute the objective functions of the operators, defined in \cref{eq:operator_obj_tx,eq:operator_obj_pt}, and of the municipality, defined in \cref{eq:social_obj}, respectively.
Further, \texttt{MLPinference} calculates approximate responses through the learned regressor.
Overall, the model-free operations \eqref{eq:middle_operator_itr} and \eqref{eq:top_operator_itr} for operators and the municipality are desirable, especially in hierarchical scenarios where accurate modeling and decoupling of different layers are restrictive.

\begin{algorithm}[!t]
\caption{Equilibrium seeking for middle-level operators}
\label{alg:middle-equilibrium}
\small
\begin{algorithmic}[1]
\Require Network $\graph$, demand $\requestset$, initial strategies $\operatoraction^0=(\operatoraction_\operatorindex^0,\operatoraction_{-\operatorindex}^0)$
\For {\textbf{each} operator $\operatorindex$} \label{algstep:for-operator}
\For{$t \gets 0, \ldots, T$}
    \State $\flow^{t} \leftarrow \texttt{Algorithm\ref{alg:fw}}(\operatoraction_\operatorindex^t \!+\! \delta v^t;\operatoraction_{-\operatorindex}^t)$
    \State $\widehat\operatorobj_\operatorindex(\operatoraction_\operatorindex^t \!\pm\! \delta v^t; \operatoraction_{-\operatorindex}^t) \leftarrow \texttt{evaluate$\operatorobj_\operatorindex$}(\operatoraction_\operatorindex^t \!\pm\! \delta v^t; \operatoraction_{-\operatorindex}^t)$
    \State Update via \eqref{eq:middle_operator_itr}
\EndFor
\EndFor
\State \Return $\operatoraction_\operatorindex^T$ \label{algstep:return_strategy}

\Statex \emph{Learn the equilibrium mapping $\mathrm{MNE}(\govaction)$ offline}

\State Sample $\{(\operatoraction^{0,i},\govaction^i)_{i=1}^{N}\} \subset \operatoractionset \times \govactionset$
\State Obtain $\operatoraction^{T,i}$ based on lines \ref{algstep:for-operator}-\ref{algstep:return_strategy}
\State Calculate $\flow^{T,i} = \texttt{Algorithm\ref{alg:fw}}(\operatoraction^{T,i}), i=1,\ldots,N$
\State Learn an MLP regressor based on the inputs $\{(\operatoraction^{T,i}, \flow^{T,i})_{i=1}^{N}\}$ and the outputs $\{(\govaction^{T,i})_{i=1}^{N}\}$
\end{algorithmic}
\end{algorithm}

\begin{algorithm}[!t]
\caption{Top-level municipality optimization}
\label{alg:top-opt}
\small
\begin{algorithmic}[1]
\Require Network $\graph$, demand $\requestset$, initial policy vector $\govaction^0$
% \State \textbf{Municipality}
\For{$t \gets 0, \ldots, T$}
    \State $(\operatoraction^{t},\flow^{t}) \leftarrow \texttt{MLPinference}(\govaction^t \pm \delta v^t)$
    \State $\widehat\govobj(\govaction^t \!\pm\!\delta v^t) \leftarrow \texttt{evaluate$J$}(\operatoraction^{t},\flow^{t})$
    \State Update via \eqref{eq:top_operator_itr}
\EndFor
\State \Return $\govaction^T$
\end{algorithmic}
\end{algorithm}
\vspace{-5pt}

\section{Numerical Experiments}\label{sec:numerical}
We evaluate the proposed framework on the mobility system of Zurich, Switzerland.
The network is built from OpenStreetMap~\cite{openstreetmap}, including road, pedestrian, and tram layers, and instantiated as in \cref{sec:Multimodal Mobility Network}.
The resulting multimodal graph has 891 vertices and 4,244 edges.
Travel demand corresponds to a one-hour slice extracted from a full-day simulation calibrated with Swiss Federal Statistical Office population data~\cite{sbfsPortal}, yielding 1,928 distinct trips.

\begin{figure}[tb]
    \centering
    \includegraphics[width=1\linewidth]{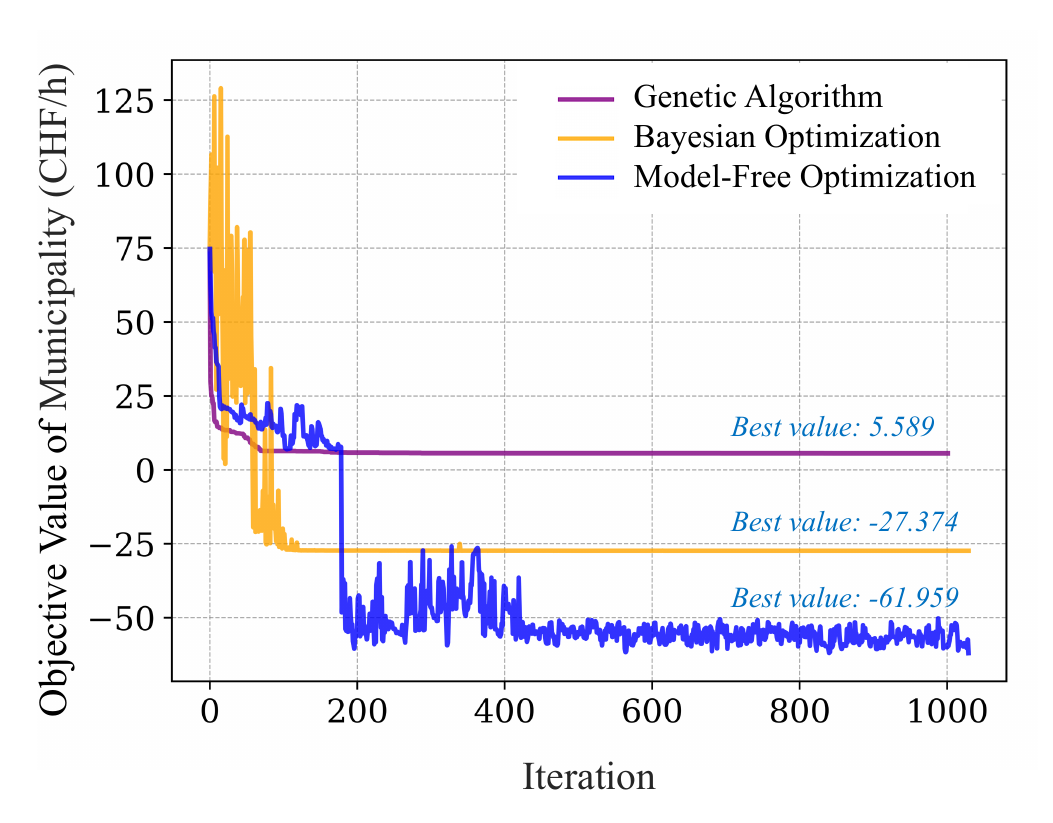}
    \caption{Evolution of the municipal objective (CHF/h) across iterations. Lower is better.}
    \label{fig:optimization process}
\end{figure}

\begin{figure}
    \centering
    \includegraphics[width=1\linewidth]{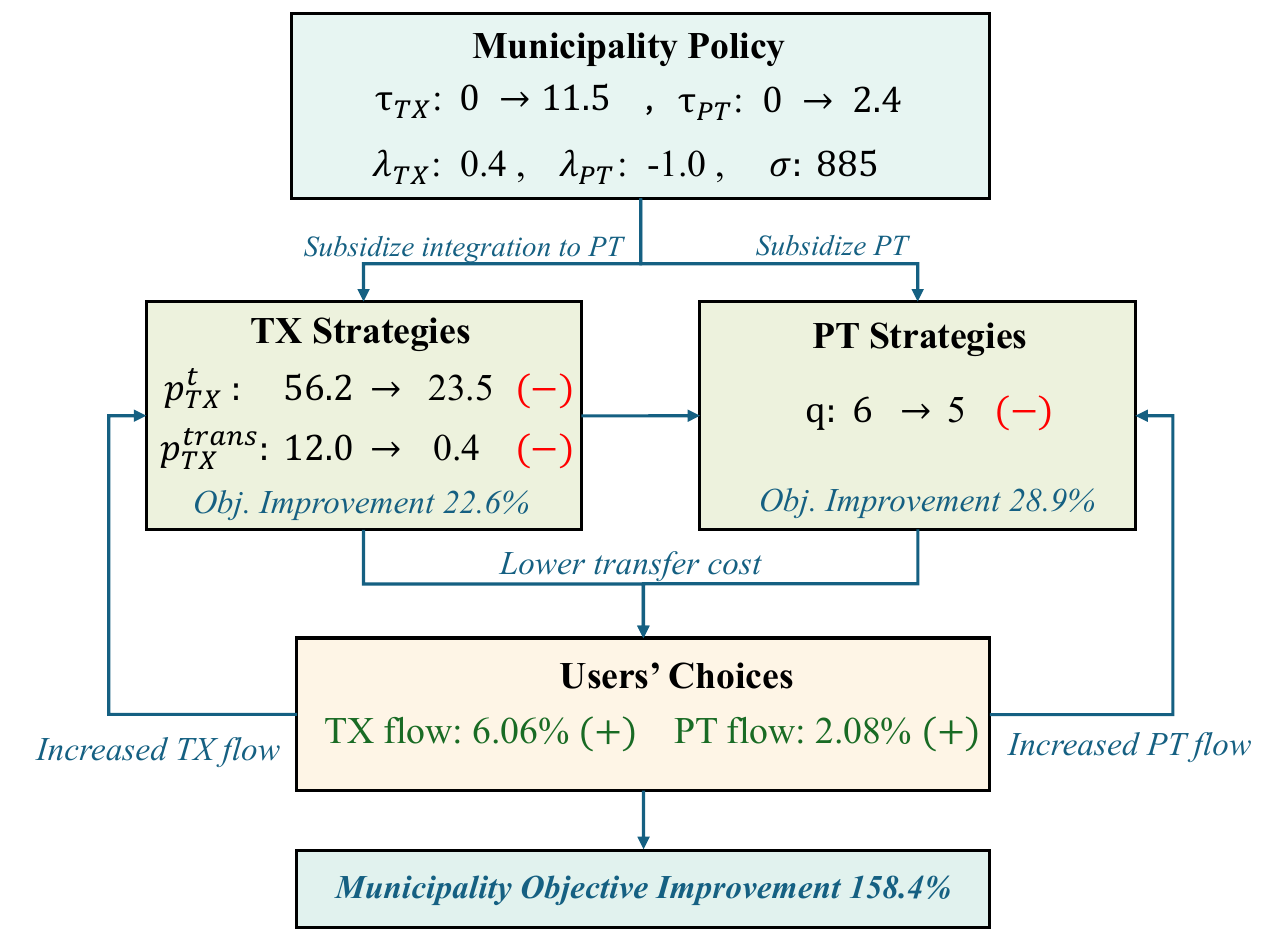}
    \caption{Impact of integration incentives on costs, transfers, and ridership composition.}
    \label{fig:integration_impact}
\end{figure}

\paragraph*{Protocol and Units}
All results are reported at equilibrium: for each municipal policy, operator strategies are updated to a middle-level equilibrium and traveler flows to a low-level equilibrium.
The municipality minimizes the social objective~$\govobj$, expressed in CHF per hour (CHF/h), reflecting the one-hour demand slice with monetary valuation of time and costs;
lower (more negative) values indicate better outcomes for the municipality.

\paragraph*{Optimization Performance}
We compare our model-free feedback method against Bayesian Optimization (BO)~\cite{shahriari2015taking} and Genetic Algorithm (GA)~\cite{katoch2021review}.
All three methods start from the same initial municipal policy and are allocated the same evaluation budget.
For BO, the best configuration uses the Expected Improvement acquisition function with the hyperparameter $\xi=0.01$ and the kernel involving Matérn, WhiteKernel, and ConstantKernel components. 
For GA, we tested population sizes in $[20,200]$, mutation rates in $[0.01,0.1]$, and crossover rates in $[0.7,0.9]$, with the best setting of population size $200$, mutation rate $0.1$, and crossover rate $0.85$. 
For the proposed method, we vary the step size $\eta$ and the exploration parameter $\delta$, obtaining the best performance with $\eta=0.018$ and $\delta=2$.

%
% \zhmargin{I feel we should provide the detailed parameters in the optimal configurations}
\Cref{fig:optimization process} shows the evolution of the municipal objective for the best-performing configuration of each method.
Our method attains a substantially better final value than both baselines.
In particular, it reaches a best objective of \mbox{$-62$~CHF/h}, corresponding to a large improvement over the initial policy.
BO plateaus at \mbox{$-27$~CHF/h} and GA remains suboptimal at \mbox{$5.6$~CHF/h}.
Empirically, BO and GA exhibit fast initial descent but stagnate as dimensionality and constraints grow, whereas the two-point updates continue to exploit structure by (i) enforcing feasibility via projections and (ii) tracking lower-level equilibria with warm starts.

% The experimental results in \cref{fig:optimization process} demonstrate that the proposed method achieves a substantially better final solution compared to the other two methods. 
% %
% Specifically, the best objective value of -62 CHF/h is obtained by the model-free method, corresponding to an improvement of 136\% from the initial value. 
% In comparison, BO attains a best value of -27 CHF/h around 400 iterations with an improvement of 102\%, while GA reached a suboptimal solution of 5.6 with an improvement of 68\%.
% \zhmargin{BO approaches the optima around iteration 200?}
% \zhmargin{explain the unit CHF/h?}
% %
% These results indicate that both BO and GA exhibit fast initial descent but soon stagnate due to their limited ability to explore and exploit the search space effectively in high-dimensional environments.

%
%
\begin{figure*}[tb]
    \centering
    \includegraphics[width=0.80\linewidth]{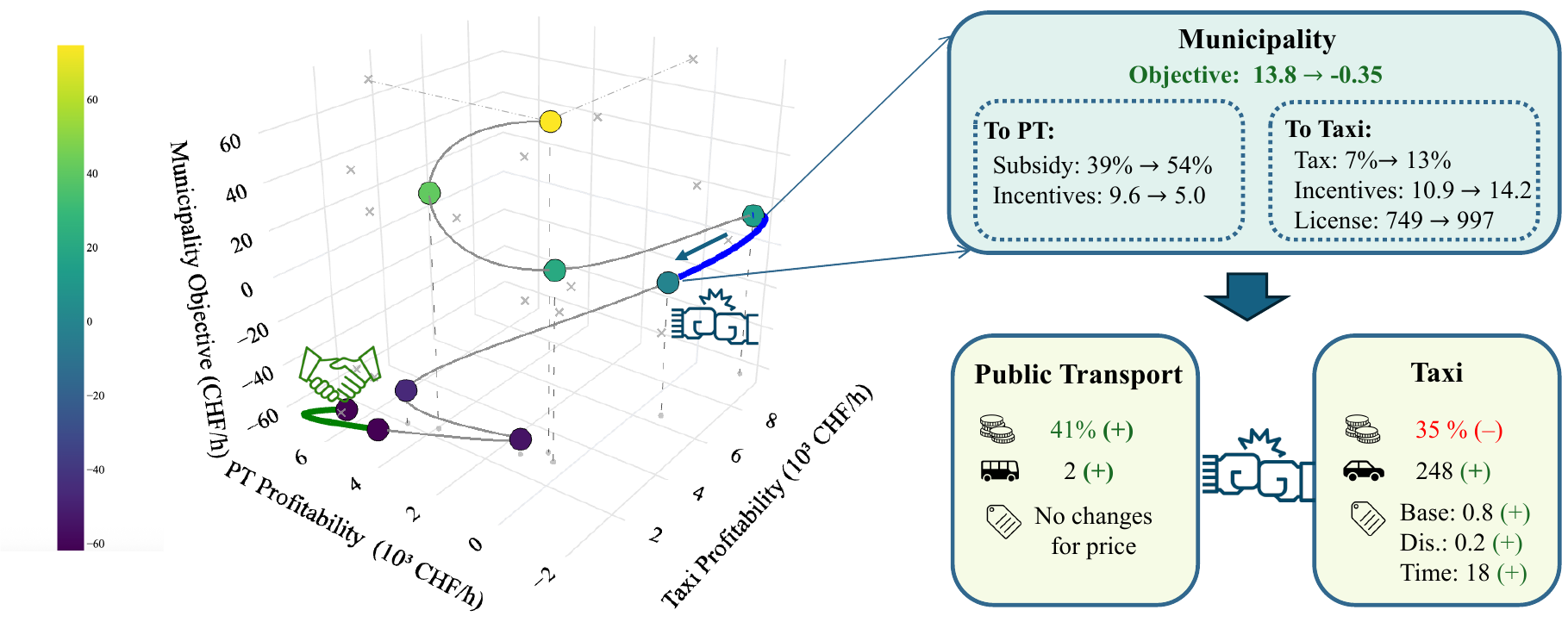}
    \caption{Policy-induced strategic interactions between PT and TX. Each point is an equilibrium under a municipal policy; color indicates the municipal objective (lower is better).}
    \vspace{-2ex}
    \label{fig:strategic interactions}
\end{figure*}

\paragraph*{Effect of Integration Incentives}
We study service-integration incentives~$\subsidy_{\pt}$ and~$\subsidy_{\taxi}$ that reward operators when a trip transfers from another service.
We compare the optimized policy against a baseline with zero incentives.
The policy found by our method is:~$\subsidy_{\pt}=2.4~\text{CHF/pass}$, 
$\subsidy_{\taxi}=11.5~\text{CHF/pass}$, 
$\tax_{\pt}=-1$, 
$\tax_{\taxi}=0.4$, 
$\license=885~\text{vehicles}$.
As shown in \cref{fig:integration_impact}, the incentives induce stronger TX--PT integration.
The taxi operator lowers prices, reducing time-based generalized travel costs from \mbox{$56.2$} to \mbox{$23.5$~CHF/h} (about a \mbox{$58\%$} decrease) and transfer costs drop to \mbox{$0.4$~CHF/pass}.
The PT operator slightly reduces frequency (from 6 to 5~trams/h), yet total PT and TX ridership both increase: lower transfer frictions make multimodal trips more attractive to previous walkers, and cheaper TX fares attract additional users.
Despite the frequency reduction, PT ridership rises due to better first/last-mile connections and lower TX access costs, which also reduce PT operating costs.
Overall, the municipality's objective improves markedly, while operators benefit as well: the TX operator's objective increases by \mbox{$22.6\%$} and the PT operator's by \mbox{$28.9\%$}. Detailed information on system performance is listed in \Cref{tab:integration_incentives}. 
These results indicate that well-designed integration incentives can align municipal, operator, and user interests.

\paragraph*{Policy-induced Operator Interactions}
Municipal policies reshape the strategic interaction between operators.
\cref{fig:strategic interactions} aggregates equilibrium outcomes encountered during policy optimization.
Each point corresponds to the joint equilibrium under a given municipal policy;
the axes report the profitability of operators, and color the municipal objective.
The blue path highlights a \emph{redistributive} policy configuration (higher PT subsidies, higher taxi taxes, looser license cap): the equilibrium shifts in favor of PT (higher profitability and frequency), while TX profitability falls by nearly 35\%, despite a larger operating fleet and higher service fees.
%with a smaller fleet and lower fares to retain demand.
By contrast, the green path illustrates a \emph{co-improving} configuration, where both operators’ objectives increase alongside municipal welfare.
These trajectories emphasize that municipal design must account for how incentives alter the game between operators, sometimes inducing complementary behavior, other times intensifying competition.

\section{Conclusion}
We presented a unified framework for hierarchical mobility systems that captures horizontal (traveler-traveler, operator-operator) and vertical (traveler-operator-municipality) interactions, and a model-free, feedback optimization scheme for municipal policy design.
The method iteratively adjusts taxes, subsidies, and operational constraints using projected two-point updates while tracking lower-level equilibria.
On Zurich’s multimodal network, our approach achieved substantially lower municipal objectives than BO and GA. 
Service-integration incentives increased multimodal usage and improved both operator objectives, illustrating how policy can steer competition toward cooperative outcomes.
Overall, our feedback optimization approach offers a practical, data-driven mechanism for governing complex mobility ecosystems.
Future work will develop formal convergence and stability guarantees under inexact equilibria and delays, handle discrete decisions with feasibility recovery, and evaluate equity-aware objectives across multiple cities.

%%%%%%%%%%%%%%%%%%%%%%%%%%%%%%%%%%%%%%%%%%%%%%%%%%%%%%%%%%%%%%%%%%%%%%%%%%%%%%%%
\section*{Appendix}
\subsection{Model Parameters}
Table \ref{appendix: model parameters} lists parameters for travelers, operators, and the Zurich municipality. [C, B, L] denote commuting, business, and leisure travelers. ‘Max.’ indicates the upper bound, with service prices capped at twice current Zurich levels.
\begin{table}[tb]
    \centering
    \caption{Overview of model parameters}
    \vspace{-2.5ex}
    \label{appendix: model parameters}
    \resizebox{0.40\textwidth}{!}{%
    \begin{tabular}{lllllll}
        \hline
        \textbf{Para.} & \textbf{Description} & \multicolumn{2}{c}{\textbf{Value}} & \textbf{Unit} 
        & \textbf{Ref.} \\ 
        \hline

        \multicolumn{6}{l}{\textbf{Low-level Model: Travelers}} \\ 
        $\vot$ & Value of time [C, B, L] & \multicolumn{2}{c}{[19, 32, 12]} & CHF/h & \cite{Axhausen2006SwissVTT} \\ 

        \multicolumn{2}{l}{\textbf{Middle-level Model: Operators}} & \textbf{Taxi} & \textbf{PT} & ~ &~ \\ 
        $P^{\mathrm{base}}$ & Max. base price & 12 & 9.2 & CHF & \cite{ZurichTaxiPrice2025,ZVVSingleTickets2025} \\ 
        $P^{\mathrm{d}}$ & Max. distance-based price & 7.6 & 5 & CHF/km & \cite{ZurichTaxiPrice2025} \\ 
        $P^{\mathrm{t}}$ & Max. time-based price & 138 & - & CHF/h & \cite{ZurichTaxiPrice2025} \\ 
        $P^{\mathrm{trans}}$ & Max. transfer price & 20 & 20 & CHF & \cite{ZurichTaxiPrice2025} \\ 
        $\ptfre_\mathrm{max}$ &  Max. service frequency & - & 16 & veh/h & \cite{VBZMetronom2022} \\ 
        $\costEnergy$ & Distance-related cost & 0.12 & 1.3 & CHF/km & \cite{CH_Gasoline_2025,en16196881,CH_Electricity_2025} \\ 
        $\costLabour$ & Time-related cost & 24 & 26 & CHF/h & \cite{TaxiDriver, TramDriver} \\ 
        $\costVehicle$ & Vehicle-related cost & 9 & 115 & CHF/veh/h & \cite{Boesch2018,VBZ_cost} \\ 

        \multicolumn{2}{l}{\textbf{Top-level Model: Municipality}} &  \textbf{Taxi}&\textbf{PT}  & \\ 
        $\tax_\mathrm{max}$ & Max. tax on revenue & 0.5 & 0.5 & - & - \\ 
        $\tax_\mathrm{min}$ & Min. tax on revenue & 0 & -1 & - & - \\ 
        $\subsidy_\mathrm{max}$ & Max. subsidy & 20 & 20 & CHF/trip & - \\ 
        $\license$ & Max. license & - & 1000 & veh & - \\
        \hline
    \end{tabular}%
    }
\end{table}

% \begin{table}[tb]
%     \centering
%     \caption{Effects of Integration Incentives }
%     \label{appendix: integration incentives}
%     \resizebox{0.32\textwidth}{!}{
%     \begin{tabular}{llll}
%         \hline
%         \textbf{Variables} & \textbf{Case 1} & \textbf{Case 2} & \textbf{Unit} \\ 
%         \hline
%         \multicolumn{4}{l}{\textbf{Municipality}} \\ 
%         $(\subsidy_{\pt}, \subsidy_{\taxi})$ & (0,0)& (11.5,2.4) & CHF/pass\\
%          $(\tax_{\pt}, \tax_{\taxi})$ & (-1,0.4)& (-1,0.4) & - \\
%          $\license$& 885 & 885 &veh\\
%          $\govobj$& 163 & -62 & CHF/pass \\
%         \multicolumn{4}{l}{\textbf{Operators}} \\ 
%         $\operatorobj_{\pt}$ & -4783 & -6166 & CHF/h\\ 
%         % $\ptfre$&
%         % $p_{\pt}^{d}$&
%         % $p_{\pt}^{\text{trans}}$&
%         % \multicolumn{4}{l}{\textbf{Taxi Operator}} \\ 
%         $\operatorobj_{\taxi}$ & -883 & -1083  & CHF/h
%         % $w$& 101
%         % $p_{\taxi}^{\text{base}}$& 
%         % $p_{\taxi}^d$&
%         % $p_{\taxi}^t$&
%         % $p_{\taxi}^{\text{trans}}$&
%         \multicolumn{4}{l}{\textbf{Flow}} \\ 
%         $y_{\pt}$& 796 & 813 & pass/h\\
%         $y_{\taxi}$& 248 & 264 & pass/h\\
%         \hline
%     \end{tabular}
%     }
% \end{table}

\cref{tab:integration_incentives} provides quantitative results on the effects of integration incentives, which are also illustrated in \cref{fig:integration_impact}.

\begin{table}[tb]
    \centering
    \vspace{-1ex}
    \caption{Effects of Integration Incentives}
    \vspace{-2.5ex}
    \label{tab:integration_incentives}
    \resizebox{0.40\textwidth}{!}{
    \begin{tabular}{lllll}
        \hline
        \textbf{Category} & \textbf{Variable} & \textbf{Case 1} & \textbf{Case 2} & \textbf{Unit} \\ 
        \hline
        Municipality & $(\subsidy_{\pt}, \subsidy_{\taxi})$ & (0, 0) & (2.4, 11.5) & CHF/pass \\
        ~ & $(\tax_{\pt}, \tax_{\taxi})$ & (-1, 0.4) & (-1, 0.4) & -- \\
        ~ & $\license$ & 885 & 885 & veh \\
        ~ & $\govobj$ & 106 & -62 & CHF/pass \\[2pt]
        Operators & $(\operatorobj_{\pt}, \operatorobj_{\taxi})$ & (-4783, -883) & (-6166, -1083) & CHF/h \\[2pt]
        Flow & $(y_{\pt}, y_{\taxi})$ & (796, 248) & (813, 264) & pass/h \\
        \hline
    \end{tabular}
    }
    \vspace{-2ex}
\end{table}

\balance
\bibliographystyle{IEEEtran}
\bibliography{reference}

\end{document}